\documentclass[a4paper,11pt]{article}
\usepackage{pos}
\usepackage{axodraw2}

\newcommand{\la}[1]{\label{#1}}
\newcommand{\ba}{\begin{eqnarray}}
\newcommand{\ea}{\end{eqnarray}}
\newcommand{\rmi}[1]{{\mbox{\scriptsize #1}}}
\newcommand{\fig}{Fig.~}
\newcommand{\eq}{Eq.~}
\newcommand{\se}{Sec.~}
\newcommand{\eqs}{Eqs.~}
\renewcommand{\nr}[1]{(\ref{#1})}
\newcommand{\tr}{{\rm Tr\,}}
\newcommand{\Tint}[1]{{\hbox{$\sum$}\!\!\!\!\!\!\!\int\,}_{\!\!\!\!\!\!\raise-0.2ex\hbox{$\scriptstyle{#1}$}}}
\newcommand{\ep}{\varepsilon}
\newcommand{\floor}[1]{\lfloor #1 \rfloor}
\newcommand{\ccdot}{\!\cdot\!}
\newcommand{\ghost}{\mbox{ghost}}
\newcommand{\gh}{\mbox{gh}}
\newcommand{\ia}{\,I_1^0}
\newcommand{\ib}{\,I_2^0}
\newcommand{\ic}{\,I_3^0}
\newcommand{\iab}{\,I_1^2}
\newcommand{\pic}[1]{\;\parbox[c]{30pt}{\begin{picture}(30,30)(0,0)
\SetWidth{1.0}\SetScale{1.0} #1 \end{picture}}\;}

\newcommand{\picb}[1]{\;\parbox[c]{45pt}{\begin{picture}(45,30)(0,0)
\SetWidth{1.0}\SetScale{1.0} #1 \end{picture}}\;}
\newcommand{\picc}[1]{\;\parbox[c]{60pt}{\begin{picture}(60,30)(0,0)
\SetWidth{1.0}\SetScale{1.0} #1 \end{picture}}\;}

\newcommand{\sbx}{\scalebox{0.725}}
\newcommand{\sby}{\scalebox{0.64}}
\newcommand{\sbz}{\scalebox{0.7}}
\def\Lwidth{1}

\def\Agl(#1,#2)(#3,#4,#5){\PhotonArc(#1,#2)(#3,#4,#5){\Lwidth}
{6.283 #3 mul 360 div #4 #5 sub #4 #5 sub mul sqrt mul Ldensity mul}}
\def\Lgl(#1,#2)(#3,#4){\Photon(#1,#2)(#3,#4){\Lwidth}
{#1 #3 sub #1 #3 sub mul #2 #4 sub #2 #4 sub mul add sqrt Ldensity mul}}
\def\Asc(#1,#2)(#3,#4,#5){\CArc(#1,#2)(#3,#4,#5)}
\def\Lsc(#1,#2)(#3,#4){\Line(#1,#2)(#3,#4)}
\def\TopoVRoi(#1){\;\pic{#1(15,15)(15,0,180) #1(15,15)(15,180,360)%
 \GCirc(0,15){3}{0} \GBoxc(30,15)(6,6){0}}\;}
\def\TopoVRooo(#1){\pic{#1(15,15)(15,-30,90) #1(15,15)(15,90,210)%
 #1(15,15)(15,210,330) \GCirc(15,30){3}{0} \GCirc(2,7.5){3}{0}% 
 \GCirc(28,7.5){3}{0}}}
\def\ToptVS(#1,#2,#3){\pic{#1(15,15)(15,0,180) #2(15,15)(15,180,360)%
 #3(30,15)(0,15)}}
\def\ToptVE(#1,#2){\picc{#1(15,15)(15,0,360) #2(45,15)(15,-180,180)}}
\def\ToprVM(#1,#2,#3,#4,#5,#6){\pic{#3(15,15)(15,-30,90) #1(15,15)(15,90,210)%
 #2(15,15)(15,210,330) #5(15,15)(2,7.5) #6(15,30)(15,15) #4(28,7.5)(15,15)}}
\def\ToprVV(#1,#2,#3,#4,#5){\!\!\picb{#2(26.25,15)(15,256,76)%
 #3(30,30)(15,30) #1(18.75,15)(15,104,284) #4(15,30)(22.5,0)%
 #5(30,30)(22.5,0)}\!\!}
\def\ToprVB(#1,#2,#3,#4){\picb{#1(30,15)(15,-120,120) #2(30,15)(15,120,240)%
 #3(15,15)(15,60,300) #4(15,15)(15,-60,60)}}
\def\TopfVX(#1,#2,#3,#4,#5,#6,#7,#8,#9){\picb{#1(15,15)(15,90,270)%
 #2(30,15)(15,-90,90) #4(30,30)(15,30) #3(15,0)(30,0) #6(15,0)(15,15)%
 #5(15,15)(30,30) #8(15,30)(20,25) #8(25,20)(30,15) #7(30,15)(30,0)%
 #9(15,15)(30,15)}}
\def\TopfVH(#1,#2,#3,#4,#5,#6,#7,#8,#9){\picb{#1(15,15)(15,90,270)%
 #2(30,15)(15,-90,90) #4(30,30)(15,30) #3(15,0)(30,0) #6(15,0)(15,15)%
 #5(15,15)(15,30) #8(30,30)(30,15) #7(30,15)(30,0) #9(15,15)(30,15)}}
\def\TopfVW(#1,#2,#3,#4,#5,#6,#7,#8){\pic{#1(15,15)(15,90,180)%
 #3(15,15)(15,180,270) #2(15,15)(15,270,360) #4(15,15)(15,0,90)%
 #5(15,15)(15,30) #7(15,15)(15,0) #6(0,15)(15,15) #8(30,15)(15,15)}}
\def\TopfVV(#1,#2,#3,#4,#5,#6,#7,#8){\!\!\picb{#2(26.25,15)(15,256,346)%
 #3(26.25,15)(15,-14,76) #4(30,30)(15,30) #1(18.75,15)(15,104,284)%
 #7(22.5,0)(15,30) #6(30,30)(26.25,15) #8(26.25,15)(22.5,0)%
 #5(25.25,15)(39.8,11.4)}\!\!}
\def\TopfVB(#1,#2,#3,#4,#5,#6,#7){\picb{#2(30,15)(15,-120,120)%
 #6(30,15)(15,120,180) #5(30,15)(15,180,240) #1(15,15)(15,60,300)%
 #4(15,15)(15,-60,0) #3(15,15)(15,0,60) #7(30,15)(15,15)}}
\def\TopfVN(#1,#2,#3,#4,#5,#6,#7){\picb{#1(15,15)(15,90,270)%
 #2(30,15)(15,-90,90) #4(30,30)(15,30) #3(15,0)(30,0)% 
 #5(15,0)(15,30) #6(30,30)(30,0) #7(15,30)(30,0)}} 
\def\TopfVU(#1,#2,#3,#4,#5,#6,#7){\pic{#3(15,15)(15,0,90)%
 #2(15,15)(15,90,180) #4(15,15)(15,180,270) #1(15,15)(15,270,360)%
 #6(0,15)(15,30) #7(15,0)(0,15) #5(30,15)(15,0)}}
\def\TopfVT(#1,#2,#3,#4,#5,#6){\pic{#1(15,15)(15,90,210)%
 #2(15,15)(15,210,330) #3(15,15)(15,-30,90) #4(2,7.5)(15,30)%
 #6(28,7.5)(2,7.5) #5(15,30)(28,7.5)}}
\def\TopfVMT(#1,#2,#3,#4,#5,#6,#7,#8){\picb{%
#1(15,15)(15,0,90)%
#2(15,15)(15,90,180)%
#3(15,15)(15,180,270)%
#4(15,15)(15,270,360)%
#5(15,30)(15,15)%
#6(0,15)(15,15)%
#7(15,15)(15,0)%
#8(37.5,15)(7.5,-180,180)}}
\def\TopfVMB(#1,#2,#3,#4,#5,#6,#7,#8,#9){\picb{%
#1(15,30)(30,30)%
#2(15,15)(15,90,180)%
#3(15,15)(15,180,270)%
#4(15,0)(30,0)%
#5(15,30)(15,15)%
#6(0,15)(15,15)%
#7(15,15)(15,0)%
#8(30,30)(30,0)%
#9(30,15)(15,-90,90)}}
\def\TopfVVTa(#1,#2,#3,#4,#5,#6,#7){\picb{%
#1(15,15)(15,0,60)%
#2(15,15)(15,60,180)%
#3(15,15)(15,180,300)%
#4(15,15)(15,300,360)%
#5(0,15)(22.5,28)%
#6(0,15)(22.5,2)%
#7(37.5,15)(7.5,-180,180)}}
\def\TopfVVBa(#1,#2,#3,#4,#5,#6,#7,#8){\picb{%
#1(15,30)(30,30)%
#2(15,15)(15,90,180)%
#3(15,15)(15,180,270)%
#4(15,0)(30,0)%
#5(0,30)(15,-90,0)%
#6(0,0)(15,0,90)%
#7(30,30)(30,0)%
#8(30,15)(15,-90,90)}}
\def\TopfVVTb(#1,#2,#3,#4,#5,#6,#7){\!\!\picc{%
#1(26.25,15)(15,-104,-5)%
#4(26.25,15)(15,-5,76)%
#2(15,30)(30,30)%
#3(18.75,15)(15,104,284)% 
#5(15,30)(22.5,0)%
#6(30,30)(22.5,0)%
#7(49,12)(7.5,-180,180)}\!\!}
\def\TopfVVBb(#1,#2,#3,#4,#5,#6,#7,#8){\picc{%
#1(30,30)(45,30)%
#4(22.5,0)(45,0)%
#2(15,30)(30,30)%
#3(18.75,15)(15,104,284)% 
#5(15,30)(22.5,0)%
#6(30,30)(22.5,0)%
#7(45,30)(45,0)%
#8(45,15)(15,-90,90)}}
\def\TopfVBT(#1,#2,#3,#4,#5,#6){\picc{%
#1(15,15)(15,60,300)%
#2(30,15)(15,120,240)%
#3(15,15)(15,-60,60)%
#4(30,15)(15,0,120)%
#5(30,15)(15,-120,0)%
#6(52.5,15)(7.5,-180,180)}}
\def\TopfVBBa(#1,#2,#3,#4,#5,#6,#7){\picc{%
#1(15,15)(15,60,300)%
#2(30,15)(15,120,240)%
#3(15,15)(15,-60,60)%
#4(30,15)(15,30,120)%
#5(30,15)(15,-120,-30)%
#6(30,15)(15,-30,30)%
#7(45,15)(7.5,-100,100)}}
\def\TopfVBSB(#1,#2,#3,#4,#5,#6,#7,#8,#9){\picc{%
#1(15,15)(15,90,270)%
#2(15,30)(15,0)%
#3(30,30)(30,0)%
#4(45,30)(45,0)%
#5(45,15)(15,-90,90)%
#6(15,30)(30,30)%
#7(15,0)(30,0)%
#8(30,30)(45,30)%
#9(30,0)(45,0)}}
\def\TopfVBST(#1,#2,#3,#4,#5,#6,#7,#8){\picc{%
#1(15,15)(15,90,270)%
#2(15,30)(15,0)%
#3(30,30)(30,0)%
#4(15,30)(30,30)%
#5(15,0)(30,0)%
#6(30,15)(15,0,90)%
#7(30,15)(15,-90,0)%
#8(52.5,15)(7.5,-180,180)}}
\def\TopfVSS(#1,#2,#3,#4,#5,#6,#7,#8){\picc{%
#1(15,15)(15,90,270)%
#2(15,30)(15,0)%
#3(15,15)(15,0,90)%
#4(15,15)(15,-90,0)%
#5(45,15)(15,90,180)%
#6(45,15)(15,180,270)%
#7(45,30)(45,0)%
#8(45,15)(15,-90,90)}}
\def\TopfVSTT(#1,#2,#3,#4,#5,#6,#7){\picc{%
#1(15,15)(15,90,270)%
#2(15,30)(15,0)%
#3(15,15)(15,0,90)%
#4(15,15)(15,-90,0)%
#5(37.5,15)(7.5,0,180)%
#6(37.5,15)(7.5,180,360)%
#7(52.5,15)(7.5,-180,180)}}
\def\TopfVTST(#1,#2,#3,#4,#5,#6,#7){\picc{%
#1(7.5,15)(7.5,0,360)%
#2(30,15)(15,90,180)%
#3(30,15)(15,180,270)%
#4(30,30)(30,0)%
#5(30,15)(15,0,90)%
#6(30,15)(15,-90,0)%
#7(52.5,15)(7.5,-180,180)}}
\def\TopfVTTTT(#1,#2,#3,#4,#5,#6){\picc{%
#1(7.5,15)(7.5,0,360)%
#2(22.5,15)(7.5,0,180)%
#3(22.5,15)(7.5,180,360)%
#4(37.5,15)(7.5,0,180)%
#5(37.5,15)(7.5,180,360)%
#6(52.5,15)(7.5,-180,180)}}
\def\TopfVBBB(#1,#2,#3,#4,#5,#6,#7,#8,#9){\picb{%
#8(20,20)(20,0,60)%
#1(20,20)(20,60,120)%
#4(20,20)(20,120,180)%
#2(20,20)(20,180,240)%
#6(20,20)(20,240,300)%
#3(20,20)(20,300,360)%
#7(20,0)(10,15,165)%
#5(2.7,30)(10,-100,50)%
#9(37.3,30)(10,130,280)}}
\def\TopfVBBT(#1,#2,#3,#4,#5,#6,#7,#8){\picb{%
#1(15,15)(15,0,60)%
#4(15,15)(15,60,120)%
#2(15,15)(15,120,240)%
#6(15,15)(15,240,300)%
#3(15,15)(15,300,360)%
#5(15,30)(10,195,345)%
#7(15,0)(10,15,165)%
#8(37.5,15)(7.5,-180,180)}}
\def\TopfVBTT(#1,#2,#3,#4,#5,#6,#7){\picb{%
#1(20,15)(15,45,135)%
#2(20,15)(15,135,240)%
#5(20,15)(15,240,300)%
#3(20,15)(15,-60,45)%
#6(20,0)(10,15,165)%
#4(5,30)(5.5,-45,315)%
#7(35,30)(5.5,-135,225)}}
\def\TopfVTTT(#1,#2,#3,#4,#5,#6){\picb{%
#1(20,15)(15,45,135)%
#2(20,15)(15,135,270)%
#3(20,15)(15,-90,45)%
#5(20,-5.5)(5.5,-270,90)%
#4(5,30)(5.5,-45,315)%
#6(35,30)(5.5,-135,225)}}
\def\TopoS(#1){\picb{#1(0,15)(20,15) #1(25,15)(45,15)%
 \GCirc(22.5,15){3}{0}}}
\def\TopoSB(#1,#2,#3){\picb{#1(0,15)(7.5,15) #2(22.5,15)(15,0,180)%
 #3(22.5,15)(15,180,360) #1(37.5,15)(45,15)}}
\def\TopoST(#1,#2){\picb{#1(0,0)(22.5,0) #1(22.5,0)(45,0)%
 #2(22.5,15)(15,-90,270)}} 
\def\ToptSi(#1){\picb{#1(0,15)(20,15) #1(25,15)(45,15)%
 \GBoxc(22.5,15)(6,6){0}}}
\def\ToptSM(#1,#2,#3,#4,#5,#6){\picb{#1(0,15)(7.5,15) #1(37.5,15)(45,15)%
 #2(22.5,15)(15,0,90) #3(22.5,15)(15,90,180) #4(22.5,15)(15,180,270)%
 #5(22.5,15)(15,270,360) #6(22.5,30)(22.5,0)}}
\def\ToptSAl(#1,#2,#3,#4,#5){\picb{#1(0,15)(7.5,15) #1(37.5,15)(45,15)%
 #2(22.5,15)(15,0,90) #3(22.5,15)(15,90,180) #4(22.5,15)(15,180,360)%
 #5(7.5,30)(15,270,360)}}
\def\ToptSAr(#1,#2,#3,#4,#5){\picb{#1(0,15)(7.5,15) #1(37.5,15)(45,15)%
 #2(22.5,15)(15,0,90) #3(22.5,15)(15,90,180) #4(22.5,15)(15,180,360)%
 #5(37.5,30)(15,180,270)}}
\def\ToptSE(#1,#2,#3,#4,#5){\picb{#1(0,15)(7.5,15) #1(37.5,15)(45,15)%
 #3(15,15)(7.5,0,180) #4(15,15)(7.5,180,360)% 
 #2(30,15)(7.5,0,180) #5(30,15)(7.5,180,360)}} 
\def\ToptSS(#1,#2,#3,#4){\picb{#1(0,15)(7.5,15) #1(37.5,15)(45,15)%
 #4(7.5,15)(37.5,15) #2(22.5,15)(15,0,180) #3(22.5,15)(15,180,360)}}
\def\TopoVRolo(#1,#2){\;\pic{#1(15,15)(15,0,180) #1(15,15)(15,180,360)%
 \GCirc(0,15){3}{0} \GCirc(30,15){3}{0} #2(15,0)(15,30)}\;}
\def\ToptVEoo(#1,#2){\;\picc{#1(15,15)(15,-180,180) #2(45,15)(15,7,367)%
 \GCirc(0,15){3}{0} \GCirc(60,15){3}{0} }\;}
\def\TopfVBBb(#1,#2,#3,#4,#5,#6,#7,#8,#9){\pic{%
#8(15,15)(15,0,60)%
#1(15,15)(15,60,120)%
#4(15,15)(15,120,180)%
#2(15,15)(15,180,240)%
#6(15,15)(15,240,300)%
#3(15,15)(15,300,360)%
#7(15,0)(7.5,15,165)%
#5(2,22.5)(7.5,-100,50)%
#9(28,22.5)(7.5,130,280)}}
\newcommand{\defDiag}[2]{\expandafter\newcommand%
  \csname diag-#1\endcsname{#2}}
\newcommand{\diag}[1]{\csname diag-#1\endcsname}
\newcommand{\picj}[1]{\;\parbox[c]{40pt}{\begin{picture}(40,40)(0,0)
\SetWidth{1.0}\SetScale{1.0} #1 \end{picture}}\;}
\newcommand{\picbj}[1]{\;\parbox[c]{60pt}{\begin{picture}(60,40)(0,0)
\SetWidth{1.0}\SetScale{1.0} #1 \end{picture}}\;}
\newcommand{\piccj}[1]{\;\parbox[c]{80pt}{\begin{picture}(80,40)(0,0)
\SetWidth{1.0}\SetScale{1.0} #1 \end{picture}}\;}
\defDiag{960}{\sbx{\picj{
\Asc(6.7,6.7)(9.3,135,315) 
\Asc(6.7,33.3)(9.3,45,225) 
\Asc(33.3,6.7)(9.3,225,45) 
\Asc(33.3,33.3)(9.3,315,135) 
\Lsc(13.3,0)(26.7,40) 
\Lsc(0,13.3)(40,26.7) 
\Lsc(0,26.7)(40,13.3) 
\Lsc(13.3,40)(26.7,0)}}}
\defDiag{992}{\sbx{\picbj{
\Asc(20,20)(20,0,360) 
\Lsc(0,20)(40,20) 
\Asc(53.3,6.7)(9.3,225,45) 
\Asc(53.3,33.3)(9.3,315,135) 
\Lsc(40,20)(46.7,40) 
\Lsc(40,20)(60,26.7) 
\Lsc(40,20)(60,13.3) 
\Lsc(40,20)(46.7,0)}}}
\defDiag{961}{\sbx{\picbj{
\Asc(20,20)(20,0,360) 
\Asc(50,20)(10,0,360) 
\Asc(20,2)(27,42,138) 
\Asc(20,38)(27,222,318) }}}
\defDiag{841}{\sbx{\picj{
\Asc(20,20)(20,0,180)
\Asc(20,30)(22.36,-153.43,-26.57)
\Lsc(0,20)(40,20)
\Asc(20,10)(22.36,26.57,153.43)
\Asc(20,20)(20,-180,0)}}}
\defDiag{1008}{\sbx{\picbj{
\Asc(20,20)(20,0,360)
\Asc(50,20)(10,0,360)
\Lsc(40,20)(4,8) 
\Lsc(40,20)(4,32)}}}
\defDiag{993}{\sbx{\!\!\picbj{
\Asc(35,20)(20,256,346)
\Asc(35,20)(20,-14,76) 
\Lsc(40,39.5)(20,39.5) 
\Asc(25,20)(20,104,284)
\Lsc(30,0)(20,39.5) 
\Lsc(40,39.5)(30,0)
\Asc(38,23)(24,136,250)}\!\!}}
\defDiag{978}{\sbx{\piccj{
\Asc(20,20)(20,0,360) 
\Asc(60,20)(20,0,360)
\Lsc(0,20)(80,20)}}}
\defDiag{952}{\sbx{\picj{
\Asc(20,20)(20,90,210)
\Asc(20,20)(20,210,330) 
\Asc(20,20)(20,-30,90) 
\Lsc(3,10)(20,40)
\Lsc(37,10)(3,10) 
\Lsc(20,40)(37,10)}}}
\defDiag{1016}{\sbx{\picj{
\Asc(20,20)(20,0,90)
\Asc(20,20)(20,-180.,-90)
\Asc(40,0)(20,90,180)
\Asc(0,40)(20,-90,0)
\Asc(0,0)(20,0,90)
\Asc(20,20)(20,-90.,0)
\Asc(20,20)(20,90,180)}}}
\defDiag{1012}{\sbx{\picbj{
\Asc(20,20)(20,0,360)
\Asc(50,20)(10,0,360)
\Lsc(40,20)(20,20)
\Lsc(20,0)(20,40)}}}
\defDiag{1010}{\sbx{\picbj{
\Asc(20,20)(20,90,270)
\Asc(40,20)(20,-90,90) 
\Lsc(40,40)(20,40) 
\Lsc(20,0)(40,0)
\Lsc(20,0)(20,40) 
\Lsc(40,40)(40,0) 
\Lsc(20,40)(40,0)}}}
\defDiag{1009}{\sbx{\picbj{
\Asc(40,20)(20,-120,120)
\Asc(40,20)(20,120,180) 
\Asc(40,20)(20,180,240) 
\Asc(20,20)(20,60,300)
\Asc(20,20)(20,-60,0)
\Asc(20,20)(20,0,60) 
\Lsc(40,20)(20,20)}}}
\defDiag{1020}{\sbx{\!\!\picbj{
\Asc(35,20)(20,256,346)
\Asc(35,20)(20,-14,76) 
\Lsc(40,39.5)(20,39.5) 
\Asc(25,20)(20,104,284)
\Lsc(30,0)(20,39.5) 
\Lsc(40,39.5)(35,20) 
\Lsc(35,20)(30,0)
\Lsc(34,20)(53,15)}\!\!}}
\defDiag{1011}{\sbx{\picj{
\Asc(20,20)(20,90,180)
\Asc(20,20)(20,180,270) 
\Asc(20,20)(20,270,360) 
\Asc(20,20)(20,0,90)
\Lsc(20,20)(20,40) 
\Lsc(20,20)(20,0) 
\Lsc(0,20)(20,20) 
\Lsc(40,20)(20,20)}}}
\defDiag{1022}{\sbx{\picbj{
\Lsc(20,0)(40,0)
\Lsc(20,40)(40,40)
\Asc(20,20)(20,-180,-90)
\Asc(40,20)(20,-90,90)
\Lsc(20,20)(40,20)
\Lsc(20,0)(20,20)
\Lsc(40,20)(40,0)
\Lsc(20,20)(20,40)
\Lsc(40,20)(40,40)
\Asc(20,20)(20,90,180)}}}
\defDiag{511}{\sbx{\picbj{
\Asc(20,20)(20,90,270)
\Asc(40,20)(20,-90,90) 
\Lsc(40,40)(20,40) 
\Lsc(20,0)(40,0) 
\Lsc(20,0)(20,20)
\Lsc(20,20)(40,40) 
\Lsc(20,40)(28,33) 
\Lsc(33,28)(40,20) 
\Lsc(40,20)(40,0)
\Lsc(20,20)(40,20)}}}
\newcommand{\TLfig}[1]{{\begin{array}{c}\diag{#1}\\[2ex] \mbox{\footnotesize #1}\end{array}}}

\title{Tackling the infamous $g^6$ term of the QCD pressure}
\author{Pablo Navarrete}
\author*{York Schr\"oder}

\affiliation{Centro de Ciencias Exactas, Facultad de Ciencias, Universidad del B\'io-B\'io\\
Avenida Andr\'es Bello 720, Chill\'an, Chile}

\emailAdd{pnavar.n@gmail.com}
\emailAdd{yschroder@ubiobio.cl}

\abstract{We report on ongoing efforts to tackle an important open problem in QCD thermodynamics,
namely an evaluation of the pressure to order $g^6$ in a weak-coupling expansion, corresponding to four loops. 
In particular, we identify a class of contributing Feynman sum-integrals with lower-loop factors,
describe the formalism to tensor decompose those, and manage to map them onto scalar master sum-integrals that
have already been evaluated in the literature.}

\FullConference{%
  Loops and Legs in Quantum Field Theory - LL2022,\\
  25-30 April, 2022\\
  Ettal, Germany
}

\begin{document}
\maketitle

%%%%%%%%%%%%%%%%%%%

\section{Introduction and Motivation}

QCD thermodynamics has been a niche, but recurrent theme at this (mainly zero-temperature) conference series,
and over the years we have witnessed useful interactions that go both ways.
For example, explorations of the number-theoretic content of perturbative expansions in quantum field theory
such as reported in \cite{Broadhurst:2016hbq} have allowed to simplify fundamental constants arising in contributions to hot QCD
from 3d lattice perturbation theory \cite{Farakos:1994xh};
while algorithmic methods that had been implemented in the finite-temperature ($T$) setting \cite{Schroder:2003kb}
have subsequently been lifted to and applied in the $T=0$ world of particle phenomenology \cite{Schroder:2005va,Schroder:2005db}.

In weak-coupling expansions within finite-temperature field theory 
one has to deal with the so-called Linde problem \cite{Linde:1980ts}, which arises from the non-perturbative nature of gluon confinement:
while strong interactions in the thermal plasma lead to an exponential screening of the color-electric part of massless 
gauge fields at large distances (in full analogy to Debye screening in an electromagnetic plasma), 
their color-magnetic part remains unscreened in perturbation theory.
In naive loop expansions this causes an infrared problem, which can be evaded within effective 
theories, as has been worked out for QCD \cite{Ginsparg:1980ef,Appelquist:1981vg} as well as the Standard Model \cite{Kajantie:1995dw}.

Interestingly, for one of the most important thermodynamic quantities such as the QCD pressure (the negative free energy; 
with phenomenological applications ranging from heavy-ion collisions via astrophysics to early-universe cosmology),
the potentially large effects of the long-range gluonic sector enter the weak-coupling expansion for the first time at four loops.
The program of how to employ a hierarchy of effective theories to consistently account for all contributions at this order, 
as sketched in \cite{Braaten:1995jr}, 
has seen major efforts towards completion, see e.g.\ \cite{Kajantie:2002wa} and references therein.
These efforts have reduced the problem to a single -- difficult but well-defined -- perturbative computation within
thermal field theory, which consists in evaluating the hard-scale four-loop contributions within a naive loop expansion
of the pressure. 

So far, results have been obtained only for scalar theory \cite{Gynther:2007bw} 
and in the large-$N_{\rmi{f}}$ limit of QCD \cite{Gynther:2009qf}.
Here, we report some technical details necessary for attacking the core of the remaining open problem, 
mainly considering pure Yang-Mills theory.

%%%%%%%%%%%%%%%%%%%

\section{Diagrammatic expansion}
\la{se:diags}

Being an equilibrium quantity,
the pressure of hot QCD can be defined in the imaginary-time formalism of thermal field theory \cite{Kapusta:1989tk,Laine:2016hma}
as the logarithm of the partition function ${Z}$ itself as
\ba \la{eq:pQCD}
p_\rmi{QCD}(T) &=& \lim_{V \to \infty} \frac{T}{V}\ln 
\int {\cal D}[A_\mu^a,\psi,\bar\psi]
\exp\Big(
-{1\over\hbar} 
\int_0^{\hbar/T} \!\!\! {\rm d}\tau \int \! {\rm d}^{d} x\, 
{L}^\rmi{E}_\rmi{QCD} 
\Big) \;,
\ea
where the path integral is over the Euclidean Lagrange density $L^\rmi{E}$ obtained after rotating $t\rightarrow i\tau$, 
and the compact temporal integration echoes the (periodic) trace ${Z}=\tr(e^{-\beta H})$ of a canonical ensemble
and leads to replacing integrals with sum-integrals in momentum space.

As motivated above, the task is to evaluate the four-loop term in a naive weak-coupling expansion of \eq\nr{eq:pQCD}, 
which in practice means that one can regulate all (UV and IR) divergences within dimensional regularization
(we use $d=3-2\ep$ and $D=4-2\ep$ throughout this note),
relying on the embedding of the calculation into the effective theory setup \cite{Braaten:1995jr,Kajantie:2002wa,Schroder:2006vz} 
in order to correctly account for 
the effects of infrared screening.

A compact and elegant method to generate all diagrams needed for the perturbative expansion of the pressure 
proceeds via a skeleton expansion (skeletons being 2-particle irreducible (2PI) graphs,
i.e.\ those that remain connected after cutting two distinct lines)
of the Schwinger-Dyson equations \cite{Kajantie:2001hv}.
While the method is generic, for the sake of illustration we will here keep fully gluonic contributions only, 
omitting all diagrams that contain ghost (and fermion) lines.

A strict loop expansion gives the perturbative series $p=p_0+\sum(p_L^{\rm 2PI}+p_L^{\rm 2PR})$,
where $p_0$ is the pressure of the free theory. 
As indicated, at each loop order $L$ the corresponding vacuum diagrams can be separated into 2PI and 2PR (2-particle reducible) types, 
where the latter are generated by the former as demonstrated in \cite{Kajantie:2001hv} and illustrated below,
hence cleanly identifying the 2PI skeletons as the key ingredient of the expansion. 

While $p_2$ and $p_3$ have been known for a long time in full QCD \cite{Shuryak:1977ut,Arnold:1994ps} 
(as has been reviewed in \cite{Nishimura:2012ee}),
the 4-loop terms $p_4^{\rm 2PI}$ and $p_4^{\rm 2PR}$ are being considered only now, 
starting from the pure Yang-Mills contributions, i.e.\ neglecting quarks by setting $N_{\rmi{f}}=0$.
We have
\ba
%% black
p_4^{\rm 2PI}&=& \la{eq:2PI}
 {1\over72} \TopfVX(\Agl,\Agl,\Lgl,\Lgl,\Lgl,\Lgl,\Lgl,\Lgl,\Lgl) %1
 +{1\over12} \TopfVH(\Agl,\Agl,\Lgl,\Lgl,\Lgl,\Lgl,\Lgl,\Lgl,\Lgl) %2
 +{1\over8} \TopfVW(\Agl,\Agl,\Agl,\Agl,\Lgl,\Lgl,\Lgl,\Lgl) %3
 +{1\over4} \TopfVV(\Agl,\Agl,\Agl,\Lgl,\Lgl,\Lgl,\Lgl,\Lgl) %4
 +{1\over8} \TopfVB(\Agl,\Agl,\Agl,\Agl,\Agl,\Agl,\Lgl) %5
\nonumber\\[3mm]&&{}
 +{1\over8} \TopfVN(\Agl,\Agl,\Lgl,\Lgl,\Lgl,\Lgl,\Lgl) %6
 +{1\over16} \TopfVU(\Agl,\Agl,\Agl,\Agl,\Lgl,\Lgl,\Lgl) %7
 +{1\over48} \TopfVT(\Agl,\Agl,\Agl,\Lgl,\Lgl,\Lgl) %8
 +\ghost \;, \\[5mm]
p_4^{\rm 2PR}&=& \la{eq:rings}
\frac16\TopoVRooo(\Agl)
+\frac12\TopoVRoi(\Agl)
+\frac14\TopoVRolo(\Agl,\Lgl)
+\frac18\ToptVEoo(\Agl,\Agl) 
+\ghost \;,
\ea
where wavy lines represent gluons.
In the 2PR part, a circle (box) on a gluon propagator represents an insertion of a 1-loop (2-loop) irreducible gluon self-energy. 
The latter can be formally obtained by cutting lines
in skeleton diagrams (regarded as functionals of the bare gluon propagator $\Delta$) at one loop higher, in all possible ways.
This leads to 
\begin{eqnarray*}
\sbz{\TopoS(\Lgl)} &=& \Pi_1^{\rm irr} \;=\; 2\delta_\Delta\,p_2^{\rm 2PI} \;=\;
2\delta_\Delta\Big\{  
 {1\over8} \sby{\ToptVE(\Agl,\Agl)}
 +\!{1\over12} \sby{\ToptVS(\Agl,\Agl,\Lgl)} 
 +\gh \Big\} 
\;=\; \frac12\sby{\TopoSB(\Lgl,\Agl,\Agl)}
+\!\frac12\sby{\TopoST(\Lgl,\Agl)}
+\gh , \\[4mm]
\sbz{\ToptSi(\Lgl)} &=& \Pi_2^{\rm irr} \;=\; 2\delta_\Delta\,p_3^{\rm 2PI} \;=\;
2\delta_\Delta\Big\{
 {1\over24} \sby{\ToprVM(\Agl,\Agl,\Agl,\Lgl,\Lgl,\Lgl)}
 +{1\over8} \sby{\ToprVV(\Agl,\Agl,\Lgl,\Lgl,\Lgl)}
 +{1\over48} \sby{\ToprVB(\Agl,\Agl,\Agl,\Agl)}
+\gh
\Big\} \\[3mm]&=&
\frac12\sby{\ToptSM(\Lgl,\Agl,\Agl,\Agl,\Agl,\Lgl)}
+\frac12\sby{\ToptSAl(\Lgl,\Agl,\Agl,\Agl,\Agl)}
+\frac12\sby{\ToptSAr(\Lgl,\Agl,\Agl,\Agl,\Agl)}
+\frac14\sby{\ToptSE(\Lgl,\Agl,\Agl,\Agl,\Agl)}
+\frac16\sby{\ToptSS(\Lgl,\Agl,\Agl,\Lgl)}
+\gh \;.\qquad\mbox{}
\end{eqnarray*}
Plugging these self-energies into the 2PR diagrams of \eq\nr{eq:rings}, one therefore obtains all diagrams 
in terms of bare propagators only, with symmetry factors coming from those of the skeletons
and the combinatorics of the cuts. 
The 2PR diagrams naturally fall into two groups, 
\ba
%% blue
p_{4,a}^{\rm 2PR}&=& \la{eq:2PRa}
 {1\over48} \sby{\TopfVBBb(\Agl,\Agl,\Agl,\Agl,\Agl,\Agl,\Agl,\Agl,\Agl)} %1
 +{1\over8} \sby{\TopfVMB(\Lgl,\Agl,\Agl,\Lgl,\Lgl,\Lgl,\Lgl,\Lgl,\Agl)} %2
 +{1\over4} \sby{\TopfVVBb(\Lgl,\Lgl,\Agl,\Lgl,\Lgl,\Lgl,\Lgl,\Agl)} %3
 +{1\over16} \sby{\TopfVVBa(\Lgl,\Agl,\Agl,\Lgl,\Agl,\Agl,\Lgl,\Agl)} %4
 +{1\over24} \sby{\TopfVBBa(\Agl,\Agl,\Agl,\Agl,\Agl,\Agl,\Agl)} %5
 +{1\over16} \sby{\TopfVBSB(\Agl,\Lgl,\Lgl,\Lgl,\Agl,\Lgl,\Lgl,\Lgl,\Lgl)} %6
 +\gh \;,\\[5mm]
%% green
p_{4,b}^{\rm 2PR}&=& \la{eq:2PRb}
 {1\over16} \sby{\TopfVBBT(\Agl,\Agl,\Agl,\Agl,\Agl,\Agl,\Agl,\Agl)} %1
 +\!{1\over16} \sby{\TopfVBTT(\Agl,\Agl,\Agl,\Agl,\Agl,\Agl,\Agl)} %2
 +\!{1\over48} \sby{\TopfVTTT(\Agl,\Agl,\Agl,\Agl,\Agl,\Agl)} %3
 +\!{1\over8} \sby{\TopfVMT(\Agl,\Agl,\Agl,\Agl,\Lgl,\Lgl,\Lgl,\Agl)} %4
 +\!{1\over4} \sby{\TopfVVTb(\Agl,\Lgl,\Agl,\Agl,\Lgl,\Lgl,\Agl)} %5
 +\!{1\over16} \sby{\TopfVVTa(\Agl,\Agl,\Agl,\Agl,\Lgl,\Lgl,\Agl)} %6
 +\!{1\over24} \sby{\TopfVBT(\Agl,\Agl,\Agl,\Agl,\Agl,\Agl)} %7
\nonumber\\[4mm]&&{}
 +{1\over8} \sby{\TopfVBST(\Agl,\Lgl,\Lgl,\Lgl,\Lgl,\Agl,\Agl,\Agl)} %8
 +{1\over16} \sby{\TopfVTST(\Agl,\Agl,\Agl,\Lgl,\Agl,\Agl,\Agl)} %9
 +{1\over32} \sby{\TopfVSS(\Agl,\Lgl,\Agl,\Agl,\Agl,\Agl,\Lgl,\Agl)} %10
 +{1\over16} \sby{\TopfVSTT(\Agl,\Lgl,\Agl,\Agl,\Agl,\Agl,\Agl)} %11
 +{1\over32} \sby{\TopfVTTTT(\Agl,\Agl,\Agl,\Agl,\Agl,\Agl)} %12
 +\gh \;,
\ea
where $p_{4,b}^{\rm 2PR}$ contains factors of lower-loop vacuum integrals that shall be our main focus here.

%%%%%%%%%%%%%%%%%%%

\section{Integral family}

To analyze the momentum-space vacuum Feynman (sum-) integrals that contribute to the evaluation of the 4-loop pressure given by the diagrams
of \eqs\nr{eq:2PI}, \nr{eq:2PRa} and \nr{eq:2PRb} in a coherent framework, 
we employ a single 4-loop family of massless inverse propagators $D_n=P_n\ccdot P_n$
constructed from the ten linear combinations $P_n$ of the loop momenta $P_{1\dots4}$ given by
\ba \la{eq:family}
\{P_1,\dots,P_{10}\} = \{P_1,P_2,P_3,P_4,P_1-P_4,P_2-P_4,P_3-P_4,P_1-P_2,P_1-P_3,P_1-P_2-P_3\} \;.
\ea
This allows to express any scalar 4-loop vacuum Feynman integral, as needed for the diagrams shown in \se\ref{se:diags},
as a list of the corresponding 10 propagator indices, where positive (negative) indices correspond to propagators (numerators).
A typical contribution to the pressure is
\ba \la{eq:1008}
\Tint{P_1}\dots\Tint{P_4} \frac{D_9D_{10}}{D_1D_2D_3[D_4]^3D_5D_6} &\equiv& I(1,1,1,3,1,1,0,0,-1,-1) \;.
\ea

Within the set of all scalar 4-loop vacuum integrals (see \fig\ref{fig:sectors} for our choice of representatives), 
we have six sectors whose denominator structure factorizes into lower-loop graphs, as indicated in the figure caption.
All diagrams of \eq\nr{eq:2PRb} can e.g.\ easily be identified with those factorizable cases; 
in fact, \eq\nr{eq:1008} is an example from sector 1008.
In general, however, the lower-loop factors correspond to tensor vacuum integrals, for which we now develop 
a tensor decomposition method in order to map them onto (1- to 3-loop) scalar sum-integrals. 
The latter had already contributed to e.g.\ $p_2$ and $p_3$ 
and can hence be considered known \cite{Nishimura:2012ee}.

\begin{figure}
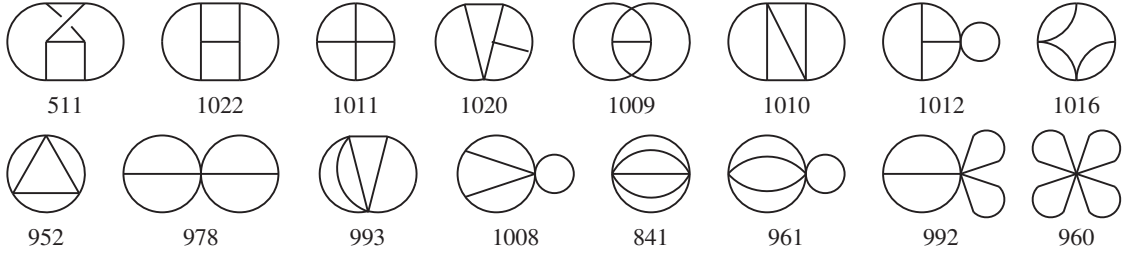

\begin{align*}
\TLfig{511}
\TLfig{1022}
\TLfig{1011}
\TLfig{1020}
\TLfig{1009}
\TLfig{1010}
\TLfig{1012}
\TLfig{1016}
\\
\TLfig{952}
\TLfig{978}
\TLfig{993}
\TLfig{1008}
\TLfig{841}
\TLfig{961}
\TLfig{992}
\TLfig{960}
\end{align*}
\vspace*{-4mm} 
\caption{\la{fig:sectors}Set of 16 unique sector representatives for the 4-loop integrals corresponding to the momentum family \eq\nr{eq:family}. 
The enumeration corresponds to the decimal representation of the respective binary footprint of propagators. Sectors 1012, 978, 1008, 961, 992 and 960
correspond to factorizable integrals.}
\end{figure}

%%%%%%%%%%%%%%%%%%%

\section{Decomposition of tensor structure}

As one marked difference with zero-temperature computations, we have 3d rotational symmetry only,
such that in tensor decompositions one has to account for an extra four-vector $U=(1,\vec 0)$.
For example, the integral of \eq\nr{eq:1008} has a 1-loop factor (line 3) in its propagator structure,
such that all numerator structures that contain the loop momentum $P_3$ should be decoupled before 
integrating the remaining 3-loop factor. 
The price to pay is to introduce zero-components $U\ccdot P=P_0$ of loop momenta $P$ into the numerator, thereby generalizing
the class of scalar sum-integrals to contain such numerators. This can easily be taken into account by adding $L$ more indices to
the list representation of an $L$\/-loop integral.

Consider an $L$-loop tensor tadpole with denominator ${\cal D}$ and (scalar part of) numerator ${\cal N}$
\ba \la{eq:TDN}
T_{\cal DN}^{\mu_1\dots\mu_n} &=& \Tint{P_1}\dots\Tint{P_L} \frac{{\cal N}(P_1,\dots,P_L,U)}{{\cal D}(P_1,\dots,P_L)}\,
\big[P_{i_1}^{\mu_1}\dots P_{i_n}^{\mu_n} 
\big]_{\mbox{\footnotesize symm in $\mu_i$}} \;.
\ea
We will deal with numerators that are symmetric in the $P^\mu$ as indicated, so
the result has to be fully symmetric in the Lorentz indices $\mu_1\dots\mu_n$, and can only depend on $U^\mu$ and $g^{\mu\nu}$. 
We use an orthogonal tensor basis (trading $g^{\mu\nu}$ for $G^{\mu\nu} = U^\mu U^\nu-g^{\mu\nu}$ with $U^\mu G^{\mu\nu}=0$)
\ba \la{eq:Ank}
A_{nk}^{\mu_1\dots\mu_n} &=& G^{\mu_1\mu_2}\cdots G^{\mu_{2k-1}\mu_{2k}}\,U^{\mu_{2k+1}}\cdots U^{\mu_n} 
\;\big|_{\mbox{\footnotesize symm in $\mu$}}
\ea
where $A_{nk} A_{nj}\sim\delta_{kj}$ (see \eqs\nr{eq:orth} and \nr{eq:pnk} below).
There are ${n\choose 2k}=\frac{n!}{(2k)!\,(n-2k)!}$ partitions of the $\mu$ into those two groups, 
and $\frac{(2k)!}{2^k\,k!}$ terms in each product $G\cdots G$,
such that $A_{nk}$ has $c_{nk}=\frac{n!}{2^k\,k!\,(n-2k)!}$ terms. 
Some examples are (abbreviating Lorentz indices as in $U^{12}\equiv U^{\mu_1}U^{\mu_2}$ etc.)
\ba
n=0 &:& A_{00}=1 \;;\nonumber\\
n=1 &:& A_{10}=U^1 \;;\nonumber\\
n=2 &:& A_{20}=U^{12} \;,\;\; A_{21}=G^{12} = U^{12}-g^{12} \;;\nonumber\\
n=3 &:& A_{30}=U^{123} \;,\;\; A_{31}=U^1G^{23} + U^2G^{31} +U^3G^{12} 
= 3U^{123}-U^1g^{23} -U^2g^{31} -U^3g^{12} \;. \nonumber
\ea
Contractions of the tensor basis elements of \eq\nr{eq:Ank} can be computed for the general-$n$ case as
\ba \la{eq:orth}
\tr(A_{nk}A_{nj}) &=& \big[A_{nk}^{\mu_1\dots\mu_n} A_{nj}^{\mu_1\dots\mu_n}\big] \;=\; \delta_{kj} \, p_{nk}\\
p_{nk} &=& \big[A_{nk}^{\mu_1\dots\mu_n} A_{nk}^{\mu_1\dots\mu_n}\big] \;=\; {n \choose 2k}\, p_k\\
p_{k} &=& \big[A_{2k,k}^{\mu_1\dots\mu_n} A_{2k,k}^{\mu_1\dots\mu_n}\big] 
\;;\quad p_0=1\;,\quad p_k=(2k-1)(D-3+2k)p_{k-1} \\
\Rightarrow p_{nk} &=& c_{nk} \,\prod_{j=0}^{k-1}(D-1+2j) \;,\la{eq:pnk}
\ea
whence all traces can be seen to reduce to simple polynomials in $D=g^{\mu\mu}=4-2\ep$.

In the tensor basis of \eq\nr{eq:Ank} any symmetric tensor tadpole of \eq\nr{eq:TDN} now decomposes as
\ba
T_{\cal DN}^{\mu_1\dots\mu_n} &=& \sum_{k=0}^{\floor{n/2}} t_{nk} \, A_{nk}^{\mu_1\dots\mu_n} \;.
\ea
To project out the coefficients $t_{nk}$, we contract with all basis tensors ($j=0..\floor{n/2}$) as
\ba
T_{\cal DN}^{\mu_1\dots\mu_n}  \, A_{nj}^{\mu_1\dots\mu_n} &=& 
 \Tint{P_1}\dots\Tint{P_L} \frac{{\cal N}(P_1,\dots,P_L,U)}{{\cal D}(P_1,\dots,P_L)}\,
 \big[P_{i_1}^{\mu_1}\dots P_{i_n}^{\mu_n} \,A_{nj}^{\mu_1\dots\mu_n} \big]
\nonumber\\
&=& \sum_{k=0}^{\floor{n/2}} t_{nk}\,\big[A_{nk}^{\mu_1\dots\mu_n}\,A_{nj}^{\mu_1\dots\mu_n}\big]
\;=\; t_{nj} \,p_{nj} \;,
\ea
where in the last step we have profited from using an orthogonal basis, such that altogether
\ba \la{eq:dec}
T_{\cal DN}^{\mu_1\dots\mu_n} &=& 
\sum_{k=0}^{\floor{n/2}} \frac{A_{nk}^{\mu_1\dots\mu_n}}{p_{nk}} \;
\Tint{P_1}\dots\Tint{P_L} \frac{{\cal N}(P_1,\dots,P_L,U)}{{\cal D}(P_1,\dots,P_L)}\,
\big[P_{i_1}^{\nu_1}\dots P_{i_n}^{\nu_n} \,A_{nk}^{\nu_1\dots\nu_n} \big] \;.
\ea

%%%%%%%%%%%%%%%%%%%

\subsection{Special case: 1-loop tensor tadpoles}

As a special case, let us apply \eq\nr{eq:dec} to 1-loop tensor tadpoles, i.e.\ $L=1$. 
Then the basis tensors are contracted fully symmetrically as 
\ba \la{eq:contract}
\big[P^{\nu_1}\dots P^{\nu_n} \,A_{nk}^{\nu_1\dots\nu_n} \big] 
= c_{nk}\,\big[UP\big]^{n-2k}\,\big[PGP\big]^k
= c_{nk} \sum_{j=0}^k {k \choose j} (-1)^{j} \big[UP\big]^{n-2j} [P^2]^j \;,
\ea
such that the most general 1-loop tensor tadpole maps onto scalar sum-integrals as
\ba \la{eq:tstDef}
T^{\mu_1\dots\mu_n}_{st} &\equiv& \Tint{P} \frac{(UP)^t}{[P^2]^s}\,P^{\mu_1}\cdots P^{\mu_n}
\;\stackrel{\tiny\nr{eq:dec}}=\;
\sum_{k=0}^{\floor{n/2}} \frac{A_{nk}^{\mu_1\dots\mu_n}}{p_{nk}} \;
\Tint{P} \frac{(UP)^t}{[P^2]^s}\,
\big[P^{\nu_1}\dots P^{\nu_n} \,A_{nk}^{\nu_1\dots\nu_n} \big]
\\&\stackrel{\tiny\nr{eq:contract}}=&
\sum_{k=0}^{\floor{n/2}} \frac{A_{nk}^{\mu_1\dots\mu_n}}{p_{nk}} \;
c_{nk} \sum_{j=0}^k {k\choose j} (-1)^j
\Tint{P} \frac{(UP)^{t+n-2j}}{[P^2]^{s-j}}
\\&=& \sum_{k=0}^{\floor{n/2}} \frac{A_{nk}^{\mu_1\dots\mu_n}}{p_{nk}}\,c_{nk} \;
\sum_{j=0}^k {k \choose j} (-1)^j\, \frac{I_{s-j}^{t+n-2j}}{I_s^{t+n}}\, I_s^{t+n}\,
\;=\; \sum_{k=0}^{\floor{n/2}} \frac{A_{nk}^{\mu_1\dots\mu_n}}{p_{nk}}\,c_{nk} \;f_{ks}\,I_s^{t+n} .\quad\mbox{}
\ea
In the last step we have used the known result for the scalar one-loop tadpole (here $d=3-2\ep$)
\ba \la{eq:i}
I_s^a &=& \Tint{P} \frac{|P_0|^a}{[P^2]^s} \;=\; \frac{2T\,\zeta(2s-a-d)}{(2\pi T)^{2s-a-d}}\,\frac{\Gamma(s-d/2)}{(4\pi)^{d/2}\,\Gamma(s)} \;,
\ea
which allowed to perform the sum over $j$ as
\ba
i_{sj} &\equiv& \frac{I_{s-j}^{t+n-2j}}{I_s^{t+n}} \;=\; 
\frac{\Gamma(s)}{\Gamma(s-j)}\,\frac{\Gamma(s-d/2-j)}{\Gamma(s-d/2)} \;=\; 
\frac{\Gamma(s)}{\Gamma(s-j)}\,\frac{\Gamma(s+1/2-D/2-j)}{\Gamma(s+1/2-D/2)} \;,\\
f_{ks} &\equiv& \sum_{j=0}^k {k \choose j} (-1)^j\,i_{sj} \;=\; 
\frac{\Gamma((D-1)/2+k)\,\Gamma((D+1)/2-s)}{\Gamma((D-1)/2)\,\Gamma((D+1)/2+k-s)} 
= \prod_{j=0}^{k-1} \frac{D-1+2j}{D+1-2s+2j} \;.\quad\mbox{}
\ea
Using \eq\nr{eq:pnk}, we observe substantial cancellations and finally obtain
\ba \la{eq:tst}
T^{\mu_1\dots\mu_n}_{st} = 
I_s^{t+n}\,
\sum_{k=0}^{\floor{n/2}} A_{nk}^{\mu_1\dots\mu_n}\,
\prod_{j=0}^{k-1} \frac1{D+1-2s+2j} \;.
\ea
Particular cases of \eq\nr{eq:tst} are
\ba
T_{st} = I_s^t \;,\quad
T_{st}^\mu = I_s^{t+1}\,U^\mu \;,\quad
T_{st}^{\mu\nu} = I_s^{t+2}\,\Big\{U^\mu U^\nu+\frac{U^\mu U^\nu-g^{\mu\nu}}{D+1-2s}\Big\} \;,\quad
\mbox{etc.}
\ea

\eq\nr{eq:tst} can now be employed to decouple some sectors into [1-loop]$\times$[3-loop], with scalar factors.
As mentioned above, owing to the appearance of $U^\mu$ in the tensor basis $A_{nk}$, this introduces factors of $P_0$ into the numerators, 
even if initially one starts from $P_0$-free sum-integrals. 

%%%%%%%%%%%%%%%%%%%

\section{Application: decoupling of sector 1008}

It turns out that, for example, the 6-line sector 1008 (cf.\ \fig\ref{fig:sectors}) contains four sum-integrals that contribute to the 4-loop pressure,
\ba
I_{10} &\equiv& I(1,1,1,3,1,1,0,0,-2,0)\;,\quad
I_{11} \;\equiv\; I(1,1,1,3,1,1,0,0,-1,-1)\;, \\
I_{12} &\equiv& I(1,1,1,3,1,1,0,0,0,-2)\;, \quad
I_{13} \;\equiv\; I(2,1,1,2,1,1,0,-1,0,-1) \;,
\ea
all of which contain scalar products in the numerator.
As an application of the above general formalism, let us demonstrate how to treat these four integrals.
Using \eq\nr{eq:tst}, we obtain the 4-loop sum-integrals in their decoupled form
\ba
I_{10}&=& \frac{4D}{D-1}\,V_4\iab -\frac4{D-1}\,L_{311}^{00}\ia\iab + L_{211}^{00}\ia\ia \;,\\
I_{11}&=& I_{10} -\frac{D}{D-1}\,X_1\iab +\frac1{D-1}\,V_2\iab  -L_{211}^{00}\ia\ia +\ic\ia\ia\ia \;,\\
I_{12}&=& 2I_{11}+J_{13}\ia-2\ic\ia\ia\ia \;,\\
I_{13}&=& X_2\ia \;,
\ea
where all sum-integrals on the right-hand sides have at most 3 loops: 
we have one-loop factors $I_s^a$ of \eq\nr{eq:i},  
two-loop factors $L_{ijk}^{00}$ (see e.g.\ App.\ B of \cite{Ghisoiu:2012yk})
as well as the specific three-loop cases\footnote{The list notation here corresponds 
to the 3-loop integral family with momenta $\{P_1,\dots,P_6\} = \{P_1,P_2,P_3,P_1-P_2,P_1-P_3,P_2-P_3\}$, with 3 indices appended to indicate 
possible powers of $\{UP_1,UP_2,UP_3\}$ in the numerator.}
\ba
J_{13} &=& I(3,1,1,1,1,-2;~0,0,0) \;, \\
V_2 &=& I(2,1,1,1,1,0;~0,0,0) \;,\quad 
V_4 \;=\; I(3,1,1,1,1,0;~0,2,0)  \;, \\
X_1 &=& I(3,1,1,1,1,0;~2,0,0) \;,\quad
X_2 \;=\; I(2,2,1,1,1,-2;~0,0,0) \;.
\ea
The first three of these have been evaluated in Refs.\ \cite{Ghisoiu:2012yk,IGthesis}.
It is in fact possible to further simplify these expressions by employing some lightweight 3-loop and 2-loop IBP reductions
along the lines of \cite{Ghisoiu:2012ph}. This allows to eliminate the $X_i$ and $L_i$ in favor of known sum-integrals, to get
\ba
I_{10}&=& \frac{4D}{D-1}\,V_4\iab +\frac{16}{(D-8)(D-3)(D-1)}\ic\ib\ia\iab -\frac1{(D-6)(D-3)}\,\ib\ib\ia\ia \;,\quad\mbox{} \\
I_{11}&=& I_{10} +\frac{D}{(D-6)(D-1)}\,V_1\iab -\frac{12-9D+D^2}{2(D-6)(D-1)}\,V_2\iab 
+\ic\ia\ia\ia 
+\nonumber\\&&{}
+\frac1{(D-6)(D-3)}\,\ib\ib\ia\ia +\frac{D}{(D-6)^2(D-3)(D-1)}\ib\ib\ib\iab \;,\\
I_{12}&=& 2I_{11}+J_{13}\ia-2\ic\ia\ia\ia \;,\\
I_{13}&=& -\frac12\,J_1\ia -J_{11}\ia -2(D-5)\,J_{12}\ia +\frac{2D-13}{D-6}\,\ib\ib\ia\ia \;,
\ea
where we have employed the additional 3-loop structures
\ba
J_1 &=& I(2,1,0,0,1,1;~0,0,0) \;,\quad
J_{11} \;=\; I(1,1,1,1,1,0;~0,0,0) \;,\\
J_{12} &=& I(2,1,1,1,1,0;~0,2,0) \;,\quad 
V_1 \;=\; I(1,2,1,1,1,0;~0,0,0) \;,
\ea
which have been evaluated in Refs.\ \cite{Gynther:2007bw,Moller:2010xw}, \cite{Andersen:2008bz,Schroder:2012hm}, 
\cite{Ghisoiu:2012kn} and \cite{Ghisoiu:2012ph}, respectively. 
We note that thus, we have reduced all four cases at hand to already known sum-integrals,
which had in fact entered earlier determinations of matching parameters within the effective theory, 
such as the Debye mass (the $J_n$ above, see \cite{Ghisoiu:2015uza} for an overview)
as well as the effective gauge coupling (the $V_n$, see \cite{Moeller:2012da,IGthesis}).

%%%%%%%%%%%%%%%%%%%

\section{Conclusions and outlook}

We have outlined techniques that allow to confront the evaluation of a specific missing four-loop contribution
to the pressure of a hot Yang-Mills gas. 
It derives its relevance from the fact that only at this order, potentially large effects from infrared logarithms that 
arise from unscreened color-magnetic gluonic modes can be accounted for in a systematic way.

Due to its technical difficulty, the problem has stood open for almost two decades now, 
and we have presented first steps towards completing the task. In particular, we have demonstrated the 
necessary tensor decomposition methods, which allow for evaluations of a class of factorizable 4-loop sum-integrals
that contribute to the pressure.

As an open problem, the remaining genuine 4-loop sum-integrals have to be evaluated up to the constant term. 
One of those can be extracted from \cite{Gynther:2007bw}, where it had contributed to the pressure of scalar theory.
Furthermore, one can explore linear IBP relations to search for simplifications. 
However, since such relations act only on the 3d part of the sum-integrals required here, they are much less powerful than at zero temperature.
 
Concerning the feasibility of the remaining evaluations of 4-loop sum-integrals, let us note that a sizable number of them
contain one-loop self-energies, for which one should be able to generalize systematic subtraction methods that have been developed 
on the 3-loop level, see \cite{Arnold:1994ps,Ghisoiu:2012yk}. 
On the other hand, evaluating sum-integrals from sectors 511, 1022 and 1011 (cf.\ \fig\ref{fig:sectors}) might represent a challenge that requires
the development of entirely new methods.

%%%%%%%%%%%%%%%%%%%

%\section*{Acknowledgments}
\acknowledgments

P.N.\ is supported by an ANID grant Mag\'ister Nacional Nr.\ 22211544; 
he is grateful to the University of Bio-Bio's postgraduate programs as well as to 
the conference organizers for facilitating his participation in LL2022.
Y.S.\ acknowledges support from ANID under FONDECYT project Nr.\ 1191073.
All figures have been prepared with Axodraw \cite{Collins:2016aya}.

%%%%%%%%%%%%%%%%%%%

\end{document}